\documentclass[acmnow]{acmtrans2m}
\usepackage{graphicx}
\usepackage{graphicx}
\usepackage{amsfonts}
\usepackage{multirow}
\usepackage{booktabs}
\usepackage{rotating}
\usepackage{tabularx}
\usepackage{enumerate}

\newdef{definition}[theorem]{Definition}
\newdef{remark}[theorem]{Remark}

\title{Solutions to Detect and Analyze Online Radicalization : A Survey}
            
\author{DENZIL CORREA and ASHISH SUREKA \\
Indraprastha Institute of Information Technology - Delhi, India.
}

\begin{abstract} 

 Online Radicalization (also called Cyber-Terrorism or Extremism or Cyber-Racism or Cyber-Hate) is widespread and has become a major and growing concern to the society, governments and law enforcement agencies around the world. Research shows that various platforms on the Internet (low barrier to publish content, allows anonymity, provides exposure to millions of users and a potential of a very quick and widespread diffusion of message) such as YouTube (a popular video sharing website), Twitter (an online micro-blogging service), Facebook (a popular social networking website), online discussion forums and blogosphere are being misused for malicious intent. Such platforms are being used to form hate groups, racist communities, spread extremist agenda, incite anger or violence, promote radicalization, recruit members and create virtual organizations and communities. Automatic detection of online radicalization is a technically challenging problem because of the vast amount of the data, unstructured and noisy user-generated content, dynamically changing content and adversary behavior. There are several solutions proposed in the literature aiming to combat and counter cyber-hate and cyber-extremism. In this survey, we review solutions to detect and analyze online radicalization. We review 40 papers published at 12 venues from June 2003 to November 2011. We present a novel classification scheme to classify these papers. We analyze these techniques, perform trend analysis, discuss limitations of existing techniques and find out research gaps.

\end{abstract}
            
         
            
            
\begin{document}
            
\maketitle

\tableofcontents

\section{Introduction}

Internet provides its users with anonymity, low publication barrier, low cost of publishing and managing content. The advent of Web 2.0 and the meteoric rise of social media has facilitated Internet users to freely disseminate ideas and opinions using multiple modalities like blogs, social networking websites, forums and video sharing websites. These characteristics are exploited by malicious online users and groups for malignant activities like hate propaganda, potential recruitment, fundraising, brainwashing \cite{JOSI:JOSI255}. The Internet is being used for various unlawful activities including a place to practice racism and xenophobia \cite{burris-smith}.

\subsection{Online Radicalization}

\emph{Online radicalization} (also called cyber extremism and cyber hate propaganda) is a growing concern to the society and also of great pertinence to governments \& law enforcement agencies.  The ease of publishing and assimilating content on the Internet via social media and video sharing websites amongst others coupled with high information diffusion rates has led to faster content dissemination and larger audience reach. This has brought together researchers around the world from various disciplines like psychology, social sciences and computer sciences to understand the problem of online radicalization and develop tools \& techniques to counter it. It has also spawned an interdisciplinary research topic : \emph{Intelligence and Security Informatics}(ISI) and a dedicated venue, IEEE International Conference on Intelligence and Security Informations \footnote{\url{http://www.isiconference.org/}} (started in 2003), where leading ISI researchers around the world meet to discuss challenges and future trends. \emph{Intelligence and security informatics (ISI) is defined as an interdisciplinary research area concerned with the study of the development and use of advanced information technologies and systems for national, international, and societal security-related applications} \cite{DBLP:conf/paisi/2011}.

Automatically detecting and analyzing online radicalization is one of the important themes of research in the domain of Intelligence and Security Informatics. Due to the inherent characteristics of the Internet, hate groups have been increasing their online presence over the years. Hate users and groups exist on various computer communication medium like WWW, Blogs, Newsgroups (Yahoo \& MSN groups). Such users also use other medium like Hosting Services, Podcasts and Games to spread xenophobic information. Various live services like Internet Relay Chat(IRC) and Internet Radio Broadcasts are available to connect with individuals to share and promote progaganda \cite{hate-dir}. The increasing presence and enormous power of social media has appealed to extremist organizations and individuals. There's a significant surge every year in the number of hate promoting groups on social networking websites like Facebook, Twitter etc \cite{simon-wiesenthal}. This makes the detection, analysis and monitoring of radical content on the Internet is a significant step in realizing a safer web. Various projects like  \emph{Dark Web} project funded by the National Science Foundation (NSF)  and the \emph{Princip} project funded by the Safer Internet Action Plan of the European Commission have sprung up in the last decade with the goal of a \emph{Safe Internet} devoid of racism and xenophobia  \cite{princip} \cite{DBLP:conf/isi/QinZLRSC05}. Law enforcement agencies and governments across the world have realized the importance of detecting and analyzing the presence of online radicalization.

\subsection{Countering Online Radicalization : Technical Challenges }

\begin{figure*}
\includegraphics[scale=0.40]{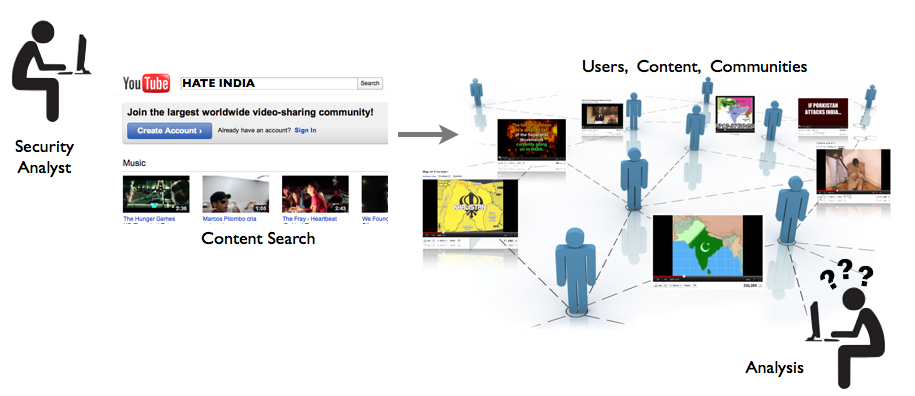}
\caption{A typical process followed by security analysts at law enforcement agencies}
\label{fig:security-analyst} 
\end{figure*}

While detecting radical online content is an important problem, there are various issues faced by security analysts working at law enforcement agencies. Such content may be extremely covert, avoiding indexing from traditional web crawlers. Much of this content also tends to be fleeting comprising of various data formats including text, image and video. The volume of the content on Internet and the mercurial rise of social media has further aggravated the challenge of discovering such content. For example, Bob is a security analyst who works for a law enforcement agency in country X. A typical working day for Bob involves finding online domestic terrorists, radical communities and content on YouTube (specifically, for his country X). YouTube, a popular video sharing website, attracts 100 million active users per week and 3 billion video views per day.\footnote{\url{http://www.youtube.com/t/press_statistics}} Bob, although a well qualified security analyst, is an elementary computer user who performs such tasks by querying the web and manually browsing through content.  Bob finds this an arduous task, particularly finding it difficult to observe common traits \& trends amongst the voluminous content. Bob also encounters problems in visualizing the data and the relationship between users who create and share such content. Figure \ref{fig:security-analyst} shows the process undertaken by security analysts like Bob working for law enforcement agencies around the world. The principal requirements of security analysts or law enforcement agent is the following \footnote{Based on inputs from senior officers from law enforcement agencies}:

\begin{enumerate}
	\item Radical online content 
	
	\item Implicit and explicit virtual communities
	
	\item Authoritative and instrumental users
\end{enumerate}

Security analysts working for various law enforcement agencies around the world empathize with Bob's situation. In addition to law enforcement agencies, various benevolent organizations like the Project for Research of Islamist Movements (PRISM) and the Search for International Terrorist Entities (SITE Institute) manually search the Internet and analyze extremist content. However, keyword based approaches may result in many false positives or off-topic content \cite{ASAP:ASAP013}. Moreover, the unstructured and informal nature of content (abbreviations, colloquialism, transliterations) adds further to the complexity of the problem. Hence, manual search for detecting, collecting and analyzing such content is not only time-consuming but also infeasible. 

In order to assist security analysts like Bob, a number of techniques have been proposed in literature to automatically detect and analyze radical content on the Internet.  The various novel techniques proposed bring multiple paradigms and perspectives to the table. This makes the task of selecting and reviewing these techniques challenging but stimulating. The first symposium of Intelligence and Security Informatics was held by National Science Foundation(NSF) / National Institute of Justice(NIJ) in 2003 at Tuscon, Arizona.\footnote{\url{http://www.isiconference.org/2003/index.htm}} Ever since, detecting and analyzing online radicalization has garnered interest from researchers across various quarters around the world. Over the past 9 years, many techniques have been proposed to tackle the problem of online radicalization detection and analysis.

\subsection{Survey : Problem Definition}

We perform an exhaustive survey on literature addressing the problem of automated online radicalization detection and its analysis. To the best of our knowledge, this is the first survey on \emph{Automated Solutions to Detect and Analyze Online Radicalization}. The solutions we survey would help security analysts to identify three attributes -- radical users, content and communities. We characterize the literature space into multiple taxonomies where each taxonomy cuts through a particular facet and has a set of associated features. These facets are carefully chosen to provide multiple perspectives on online radicalization detection approaches. We envisage these facets to be useful for both -- researchers and law enforcement agencies. Researchers can refer to this survey to understand various current approaches, while law enforcement agencies can refer to this survey to understand the state-of-the-art techniques. This survey would assist security analysts to build tools in order to detect and analyze online radicalization, keeping in mind the limitations, challenges \& issues with these techniques.

%Authorship analysis deals with the problem of identifying the true or actual owner of the content. While it helps in assigning content to actual owners, it is not a core component in a framework to detect online radicalization. Affect Analysis uses techniques to understand the sentiment or opinion of the author towards topic(s). Affect analysis helps understand the reasons why users indulge in online radical activities by giving an insight into their opinions on various topics. However, automated online radicalization detection techniques focus on detecting radical content rather than analyzing the sentiments by users. Both authorship analysis and affect analysis are used for analyzing content once detected by the automated online radicalization detection algorithms. 

\subsection{Contributions}

The main contributions of this paper are as follows :
\begin{enumerate}
	
	\item This is the \textbf{first survey} in the literature space of \emph{Automated Solutions to Detect and Analyze Online Radicalization} (also called cyber extremism, hate propaganda, cyber racism)
	
	\item We propose a novel classification scheme across multiple meaningful dimensions to help both researchers and law enforcement agencies
	
	\item We compare and contrast techniques used to detect and analyze online radicalization. We analyze these techniques to inspect trends and point out research gaps .
	
\end{enumerate}

The rest of the paper is organized as follows :  Section \ref{sec:lit-rev-process} explains the literature review process followed in selecting and analyzing papers,  Section \ref{sec:dim-article-classification} discusses the facets for paper classification, Section \ref{sec:survey-solutions-detection} and Section \ref{sec:survey-solutions-analysis} classifies the selected articles into different facets and briefly outlines the approaches, Section \ref{sec:discussion} analyzes the techniques used, trends observed and limitations of the proposed approaches in literature, Section \ref{sec:research-gaps} discusses research gaps, Section \ref{sec:conclusion} gives concluding remarks.

%For example, the \emph{Dark Web Project} funded by the National Science Foundation (NSF) collects and analyzes content by U.S. domestic, Middle Eastern, and Latin American terrorist and extremist groups.\footnote{\url{http://ai.arizona.edu/research/terror/crdabstract.asp}}\footnote{\url{http://www.nsf.gov/awardsearch/showAward.do?AwardNumber=0709338}} 

\section{Literature Review }
\label{sec:lit-rev-process}

The objectives of the work presented this study is as follows :

\begin{enumerate}[(a)]
	\item \emph{To investigate of different techniques used to detect and analyze radical content on the Internet}
	
	\item \emph{To perform an in-depth study of these techniques to analyze common trends}
	
	\item \emph{To examine limitations in these techniques and point out research gaps}
	
\end{enumerate}

\subsection{ Review Process }

\begin{figure*}
\includegraphics[scale=0.40]{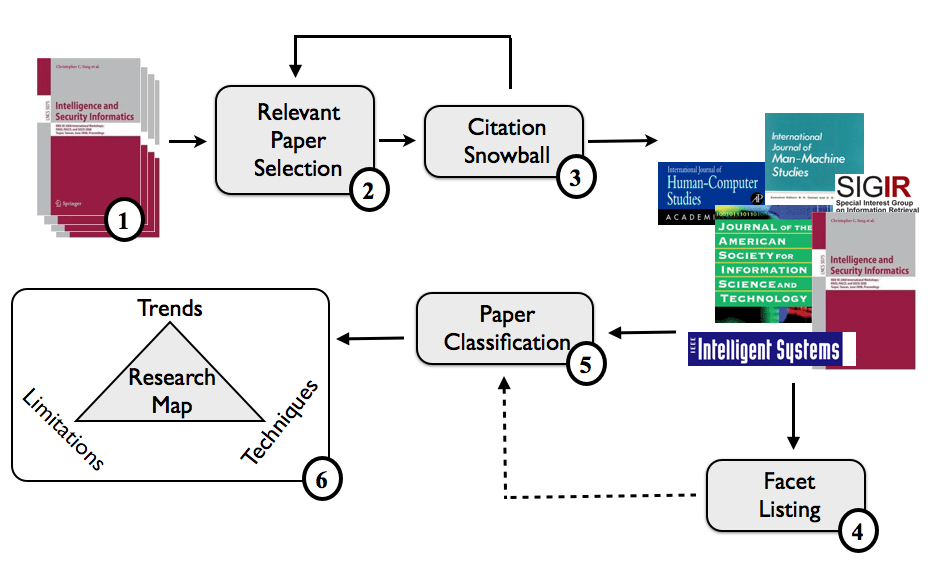}
\caption{Literature review process followed in our survey}
\label{fig:literature-review} 
\end{figure*}

In order to meet the above objectives, we collected relevant literature by following these steps. Figure \ref{fig:literature-review} shows the literature review process followed in the survey.

\begin{enumerate}
	\item \textbf{Collection} -- The first step involved collecting all papers in the IEEE Intelligence and Security Informatics(ISI) Proceedings\footnote{\url{http://www.isiconference.org/}}
	
	\item \textbf{Relevant Paper Selection} -- We then filtered out papers whose main contributions were to develop automated techniques for online radical content detection or analyze online radical content
	
	\item \textbf{Citation Snowball} -- We then performed citation analysis by browsing through the references of these papers and repeat Steps (2) and (3) till all relevant literature was exhausted
	
	\item \textbf{Facet Listing} -- In this step, we listed down various facets according to which the papers would be classified. The authors of this paper individually performed this task to come up with a paper classification scheme. The authors then sat together and finalized the facet list ironing out disagreements in the process. This ensures rigor in forming the classification scheme and reduces potential bias.
	
	\item \textbf{Paper Classification} -- This step classified each paper into its appropriate facet bucket. Just as the previous step, the authors performed the classification individually. The authors then discussed disagreements on the paper classification buckets till a convergence was achieved.
	
	\item \textbf{Solution Analysis} -- Finally, we present an analysis of the selected literature across various facets, discuss observed trends, limitations and also point out research gaps.
	
\end{enumerate}

\subsection{Related Problems and Scope}

There are several closely related research problems to automated detection and analysis of online radicalization in the field of ISI. Some of them include Deception Detection, Infrastructure Protection \& Cyber Security and Criminal Data Mining \& Network Analysis. 

\begin{itemize}

	\item[.] Deception detection is the task of detecting untruthful and subterfuge content \cite{springerlink:10.1007/978-3-540-25952-7_13}. Deceptive content generated with the objective of propaganda, brainwashing etc. may mislead Internet users. On the other hand, detecting online radicalization concerns more with discovering racist, hate promoting and xenophobic online content.
	
	\item[.]  Infrastructure protection \& Cyber security deals with the protection of critical cyber infrastructure against malicious attacks \cite{5484748}. These attacks may not necessarily be generated by radical users. 
		
	\item[.] Criminal data mining \& Network analysis concerns with mining and analyzing criminal network data to reveal hidden patterns and insights \cite{Derosa:2004:DMD:994043}. This data is generally recorded by law enforcement agencies as a formal process in their dealings with daily crime. 

\end{itemize}

In contrast to above problems, we focus on techniques which are used to detect \& analyze radical content available on the Internet. These solutions are devised to assist security analysts working for law enforcement agencies to understand unlawful usage of the Internet, in particular for radicalization. These solutions also cover analysis and summarization of online radical content, users and their online relationships. Hence, papers with  topics like Deception Detection, Infrastructure Protection \& Cyber Security and Criminal Data Mining \& Network Analysis as their main goal are not considered and beyond scope of this survey.

\subsection{Statistics}

We have rigorously chosen a collection of 40 papers published at 7 venues and 5 journals from June 2003 to November 2011. These papers were selected as they have listed down their main contribution as either a solution for automatically detecting online radicalization or performed analysis on online radical content. Table \ref{tab:paper-distr} shows the distribution of papers across conference venues and journals.

\begin{table}[h!]
\caption{Distribution of papers across conferences and journals}
\begin{center}
	\begin{tabular} {c|c}
	\toprule
		~ & \textbf{Number of Papers}  \\
		\midrule
		Conference  &  28 \\ 
		Journals & 12    \\ \hline
		\textbf{Total} & 40    \\ 
	\bottomrule
	\end{tabular}
	\label{tab:paper-distr}
\end{center}
\end{table}

\section{Facets for Article Classification}
\label{sec:dim-article-classification}

The goal of this survey is to provide a research map of the literature on online radicalization detection and analysis techniques. This would help both researchers as well was security analysts at law enforcement agencies to understand the structure of the research. In order to achieve this goal, we identify multiple facets from existing papers which would help characterize the solutions. These facets also provide key insights into trends, limitations and advantages of the techniques. 

In order to provide a better structure to the research map, we divide our survey into two parts and classify articles accordingly : \emph{Techniques for Online Radicalization Detection} and \emph{Automated techniques for Online Radicalization Analysis}.  We define the two problem as follows:

\begin{itemize}
	\item[\checkmark] \emph{Online Radicalization Detection} is the problem of detecting radical (hate promoting or extremist) content on the Internet. Detecting online radicalization is a genre-centric information retrieval problem. Typically the input is WWW (or a part of it) and the output is the detected extremist content. 
	
	\item[\checkmark] \emph{Online Radicalization Analysis} on the other hand pertains to analyzing extremist content to gain deeper insight into the functioning of extremist groups on the Internet and their usage. Typically the input to an analysis algorithm would be documents containing radical content and the output is a detailed analysis explaining the behavioral, structural and linguistic characteristics of the content. 
	
\end{itemize}

We performed a meticulous analysis of literature space and concluded that this structure would help us draw better insights and analyze trends. In addition, it provides us the flexibility to choose different facets for two closely related problems resulting in a better organization of literature. A few articles in the literature contributed to both automatic detection and analysis of online radicalization. These articles are individually surveyed with respect to both contributions and hence, included in both parts of the survey. They are also characterized according to two sets of facets in both parts of the survey.

\begin{table}[h!]
\caption{Distribution of papers surveyed which address the problem of online radicalization detection or analysis}
\begin{center}
	\begin{tabular} {c|c}
	\toprule
		\textbf{Problem} & \textbf{Number of Papers}  \\
		\midrule
		Online Radicalization Detection & 18  \\ 
		Online Radicalization Analysis & 22    \\ \hline
		\textbf{Total} & 40    \\ 
	\bottomrule
	\end{tabular}
	\label{tab:paper-stats}
\end{center}
\end{table}

We perform a rigorous evaluation of the literature in the area. We then identify properties in each article to characterize the proposed techniques. These properties are analyzed to form an initial list of facets. These facets are then discussed and refined to form a final list. Each facet was chosen to succinctly position all articles  in order to help researchers and security analysts locate techniques as per their requirements.

\subsection{Online Radicalization Detection : Facets}

In this part of the survey, we identify literature which contribute towards automatic online radicalization detection either as their main goal or a part of their goals. We search for literature whose main aim is to search, collect and detect extremist content on the Internet. For example, we collect papers whose primary objective is to find extremist websites or hate promoting online forums. Concretely, the algorithms of papers in this section of the survey receive the WWW as input and the algorithms proceed to output extremist content on WWW. We then proceed to analyze each paper and create a list of facets to create a structured research map. Each facet has a list of properties associated with it. The facets are as follows :

\begin{itemize}
	
	\item[*] \textbf{Technique} : What techniques are used to detect radical content on the web?
	
	\item[*] \textbf{Features} : What features do these techniques use?
	
	\item[*] \textbf{Modalities} : Which modalities on the web do they consider?
	
	\item[*] \textbf{Data Type} : What data types do these techniques cater to?
	
	\item[*] \textbf{Output} : How are the results presented?
	
	\item[*] \textbf{Evaluation} : How are these techniques evaluated?
	
	\item[*] \textbf{Language} : What languages do these techniques address?
	
	\item[*] \textbf{Genre} : What genre of hate is being detected?
	
\end{itemize}

These facets and it's properties along with brief descriptions are listed in Table \ref{tab:facets-detection}.

\subsubsection{Technique}

The key differentiator between 	proposed solutions for online radicalization detection. The most common technique used to detect radical content on the web is using a Link Based Bootstrapping (LBB) algorithm. There are other techniques like machine learning and multi-agents  which have also been explored for detecting racist, hate promoting content. 

\textbf{Web mining} is the process of applying data mining techniques to unearth information from web documents \cite{Etzioni:1996:WWQ:240455.240473}. Web mining can be further categorized into three parts : Web content mining, Web structure mining and Web usage mining \cite{Kosala:2000:WMR:360402.360406}. Web content mining entails the process of discovering relevant useful information from web documents including text, audio, video, metadata and hyperlinks. Web structure mining refers to the discovery of the structure of the web by exploiting the hyperlinks between web documents \cite{781636}. Web usage mining concerns with the analysis of patterns in web usage data which includes search logs, clickthrough data, server logs etc. \cite{Cooley_Mobasher_Srivastava_1997}. The above three categories are not necessary mutually exclusive and hybrid approaches have been successfully used to discover relevant information from web documents \cite{Kosala:2000:WMR:360402.360406}. Major solutions proposed to detect online radicalization employ a combination of web structure mining and web content mining \cite{DBLP:conf/isi/QinZLRSC05}. Link Based Bootstrapping approach is a type of web structure mining which harnesses the hyperlink structure of the websites or the social links between users. Typically, it starts with some seed data as input and then use the hyperlinks of websites (or social links between users) to find related content. Spiders with appropriate filtering parameters and proxies are employed to help the collect the documents while keeping in interest adversarial conditions \cite{4076513}. Other processing steps like duplicate content removal, collection update procedures are performed to maintain the integrity and consistency of data. Link Based Bootstrapping (LBB) techniques for online radicalization detection are discussed in section \ref{sec:survey-webmining-detection}.

\textbf{Text classification} (also known as text categorization) is the task of arranging text documents in pre-defined classes or categories. Machine Learning techniques are widely used for text classification \cite{Sebastiani:2002:MLA:505282.505283}. Text classification tasks follow a typical pipeline which includes : manual labeling of documents into classes, document representation, training a classifier on seen data and evaluation on an unseen test set. Documents are represented as feature vectors and these feature vectors are provided as input to train the classifier. There are various feature representations which take into consideration the content and structure of the document, if available. The simplest feature representation consists of treating each term in the document as a feature, also called Bag-of-Words (BOW). Other feature representations include tf-idf, bigrams, Part-of-Speech(POS) tags and syntax trees. Various types of classifiers like Naive Bayes, Support Vector Machines (SVMs) and Decision Trees have been used for classification \cite{Sebastiani:2002:MLA:505282.505283}. Text classification techniques are generally evaluated on four parameters : Precision, Recall, F-Measure and Accuracy.  The problem of detecting online radicalization can be treated as a binary (also called 2-class) text classification problem. Document representations include graphs, word level, character level, content, syntactic and lexical features. Machine learning classifiers like Naive Bayes, C4.5, SVM are used to train on the feature representations. Appropriate evaluation metrics like Precision, Recall, F-measure and Accuracy are reported to test the performance of the classification. Text classification techniques to detect online radicalization are discussed in section \ref{sec:survey-textclassification-detection}.

LBB and text classification are not the only techniques used to detect online radicalization. A few approaches like multi-agent based methods have also been used. We classify these methods into the \emph{other} category. Section \ref{sec:survey-other-detection} discusses these specific methods in detail.

%The idea centers around using a set of seed data which have been identified by domain extremism experts. This seed data is then expanded using hyperlinks, both in and out,  

\subsubsection{Modalities}

The Internet consists of various forms of communication modalities like websites, weblogs (also commonly known as blogs), social media (Twitter, Facebook, MySpace), image sharing websites (Flickr, Photobucket), video sharing websites (YouTube, Dailymotion) and online forums. Various terrorist organizations have created websites on the Internet \cite{ASAP:ASAP013} \cite{terror-net}. These websites help the organizations in carrying out activities like fundraising, propaganda and recruiting \cite{LEE01022002}.  Blogs (or weblogs) are online equivalent of diaries. Blogs are an effective and powerful medium to communicate ideas and reach out to other like minded users. Hence, it's a befitting medium for hate propaganda and sharing ideology \cite{Chau200757}. Online forums (or a message board) are websites designed for users to post messages, news and other content. Forums foster discussions on specific topics, increase collaboration and help organized sharing of content. In addition, there are various freely available softwares to create and host online forums. Most forums also require signing up in order to prevent detection from search engines. These characteristics of online forums appeal to terrorist organizations to organize and spread information. Studies show that many forums on the web are racist in nature \cite{Reid_Qin_Zhou_Lai_Sageman_Weimann_Chen_2005}. Social media like Facebook and Twitter have witnessed a meteoric rise in the recent past attracting a large user base. For example, Facebook has more than 800 million active users and used in more than 70 languages. \footnote{\url{https://www.facebook.com/press/info.php?statistics}} Due to enormous growth of social media, hate promoting users have turned to these sites to promulgate hate content and actively seek to connect individuals. 

\subsubsection{Data Type}

The content on the web consists of various data types like text, images, audio and video. Terrorist organizations utilize all these types to distribute material. Radicalization detection techniques focus on detecting all such variety data types. Multimedia processing is a relatively expensive task than text processing. Hate promoting videos are poor quality in nature which makes the use of multimedia difficult. Hence, most techniques rely on use of textual content or user generated content surrounding the multimedia. 

\subsubsection{Features}

Various types of features are used to assist techniques to detect online radicalization. The two commonly used features in the literature space of online radicalization are \emph{link based}  and \emph{content based} features. 

\textbf{Link based} features analyze the structure between documents to make decisions on detection of radical content. Link features include hyperlinks between web sites, replies on common threads in online forums and relationships between users on social network websites.

\textbf{Content based} features leverage the structure and content of a document. These include lexical (frequency of letters, average word length etc.), syntactic (frequency of punctuation words, function words etc.), domain specific, graph based and structural features (html markup, paragraphs etc.) Link based features are mainly used in LBB techniques while content based features are heavily used in text classification techniques. A few techniques use a combination of both link based and content based features.

\subsubsection{Evaluation}

Precision, Recall, F-Measure and Accuracy are widely used and acceptable evaluation metrics. Precision measures the ``exactness" of the technique while Recall measures the ``completeness". Accuracy, as the name suggests, evaluates the ``correctness" of the solution. F-measure is the weighted harmonic mean of precision and recall, essentially combines the two metrics into a single unit of measurement. Lets consider Table \ref{tab:evaluation-metrics} to formally define the evaluation metrics. We now formally define the evaluation metrics as follows : \\

\begin{table}[htbp]
  \centering
  \caption{Evaluation data}
    \begin{tabular}{rrrr}
    \toprule
          ~ &       & \multicolumn{2}{c}{Actual Class}  \\
             \midrule
         ~  &       & True  & False \\
              \midrule

    \multicolumn{1}{c}{\multirow{2}[0]{*}{Predicted Class}} & Positive & True Positive(TP)    & False Positive(FP) \\
    \multicolumn{1}{c}{} & Negative & False Negative(FN)    & True Negative(TN) \\
    \bottomrule
    \end{tabular}%
  \label{tab:evaluation-metrics}%
\end{table}%

 Precision = $\frac{TP}{TP + FP}$ \\

Recall = $\frac{TP}{TP + FN}$ \\

F- Measure = $\frac{2 * Precision * Recall}{Precision + Recall}$ \\

Accuracy = $\frac{TP + TN}{TP + FP + TN + FN}$ \\

An increase in precision signifies less false positives while a increase in recall signifies less false negatives.

\subsubsection{Output}

There are three categories of output which are presented : users, content and communities. Users are simply Internet users on the web, online forums and social media who identify themselves according to the screen name or a pseudonym. Content refers more to the type of documents and includes text and multimedia(images, video and audio). A common property observed amongst most structured networks  like the WWW, social networks and collaboration networks is the formation of communities. A \emph{community} can be loosely defined as a collection of individuals with common interests or ideas. Analysis of communities and their interaction can reveal important insights like central leaders, structure and information flow.

\subsubsection{Language}
\label{subsubsec:language}

Racist and hate promoting content is prevalent in multiple languages. Different languages create associated issues especially for radicalization detection techniques which utilize content. For example, Arabic text is written from right to left leading to a fundamental change in which document representations are formed from arabic text. There may be multi-lingual content (both Arabic and English) present in the document. This leads to rethinking of methods to extract structural and syntactic features.

\subsubsection{Genre}
\label{subsubsec:genre}

Various genres of xenophobia exist on the Internet. Some of them include US domestic extremism, Middle East extremism, Anti-Semitism, hate against Blacks and anti-India hate. We also classify our articles into one of these genres.


\subsection{Online Radicalization Analysis : Facets}

In this part of the survey, we identify literature which contributes towards analyzing online radicalization either as their main goal or a part of their goals. We search for literature whose main aim is to analyze extremist content on the Internet. For example, we collect papers whose primary objective is to analyze extremist websites or hate promoting online forums. Concretely, the algorithms of papers in this section of the survey receive extremist (hate promoting or racist) content as input and the algorithms proceed to analyze extremist content on various parameters. For example, extremist videos on YouTube can be analyzed to determine users posting such extremist videos and the communities these users form. Moreover, social network analysis can also determine leaders in these communities.  Once we identify the literature, we proceed to analyze each paper and create a list of facets to create a structured research map. Each facet has a list of properties associated with it. The facets are as follows :

\begin{itemize}
	
	\item[*] \textbf{Types of Analysis} : What are the types of analysis used to examine radical content on the web?
	
	\item[*] \textbf{Link Analysis} : What are the type of network analyses which are performed?

	\item[*] \textbf{Content Analysis} : What are the type of content analyses which are performed?

	\item[*] \textbf{Modalities} : Which modalities on the web do they consider?
			
	\item[*] \textbf{Language} : What languages do these techniques address?
	
	\item[*] \textbf{Genre} : What genre of hate is being detected?
	
\end{itemize}

These facets and it's properties along with brief descriptions are listed in Table \ref{tab:facets-analysis}.

\subsubsection{Type of Analysis}

There are various types of analysis performed on online radical content to gain insights and help understand the nature of the posted content. Extremist websites often link to each other and gather together to form a peculiar structure. It's important to analyze this structure to understand how these websites connect with each other and study their interactions. Web link analysis (or Web link mining) refers to the process of modeling and analyzing links between websites \cite{781636}. Link analysis can help discover communities, find central leaders and characterize various network properties. Content analysis on the other hand analyzes the content in the web sites. Linguistic patterns and textual clues can help understanding the nature of these websites and the reason behind the formation of these websites. It can also help throw light on the richness of the content and hence, depicting the technical sophistication of  extremist websites. One can also gauge the affect (or emotion) towards various topics by analyzing the content. Moreover, these patterns can also reveal the objectives behind  which these websites are created and the manner these websites are being used. In some scenarios, performing link analysis or content analysis individually is insufficient. Hence, a combination of both link and content analysis is used in such scenarios.

\subsubsection{Network Based Analysis}

Network based analysis (also called Link Analysis) is used to understand the structure of  links between hate promoting websites. The existence of a link between two nodes can be due to a hyperlink, friend relationship or due to some kind of  interaction like a reply to e-mail or bulletin message. Link analysis helps examine the topology of the network and characterize the properties of the network. Average shortest path length, clustering coefficient and degree distribution are common metrics used to reveal properties of a network. Average shortest path length is the mean of all shortest paths between every node in a network. It measures the connectivity between any two of nodes in a network. Clustering coefficient measures the probability that two nodes in a network would belong to the same community. In-degree measures the number of links in to a node while out-links measures the number of links flowing out from a node. Degree distribution is the probability distribution of node degrees over a network. A network which follows a high clustering coefficient  is termed as a small world network. A network whose degree distribution follows a power law is termed as a scale-free network \cite{wasserman_faust_94}. 

An important property of every social network is the ability of nodes in the network to group together as communities. Communities can be defined as a group of nodes with a common interest or topic. The group of nodes which form a community are said to possess more intra-community links than inter-community links. Various community detection algorithms have been proposed to identify the formation of communities \cite{Girvan:529744}  \cite{PhysRevE.69.026113} \cite{PhysRevE.69.066133}. The usage of such community detection algorithms helps understand the underlying community structure in extremist websites, forums and blogs. The interaction between these communities  are also of great interest to help understand information flow. Blockmodeling can be used to understand the interactions between communities \cite{wasserman_faust_94}. Visualization of communities is also an important issue as this information is consumed by elementary computer users. Various visualization methods like Multi Dimensional Scaling (MDS) and Snowflake Visualization have been used to understand the formation of communities \cite{freeman2000visualizing}.

A subset of nodes in a network called central nodes are the key to the existence and structure of the network. Central nodes generally act as nodes which link to most parts of the network or form ``bridges" between communities. In order to identify central nodes in a network, three well studied measures are used : degree, betweenness and closeness \cite{LintonC19781979215}. Degree measures the number of links incident on a node and thus is the measure of the node's \emph{popularity}.  The number of shortest paths passing through a node in a network gives its ``betweenness'' measure. The sum of the length of shortest paths between one node and all other nodes in a network gives its ``closeness'' measure. Nodes with high betweenness are known to act as links between communities while nodes with high closeness are well connected to most parts in the network. The removal of central nodes in a network can significantly alter the structure of the network.

\subsubsection{Content Based Analysis}

Content based analysis is used to investigate into the content posted by extremist groups on the Internet. Affect analysis is used to understand the affect (or emotion) in text towards topics. Affect analysis can help gauge the intensity of violence, racism and hatred in extremist content. Authorship analysis is used to identify the owner of a written piece of text based on the author's stylistic cues. It can be used to detect Internet users in adversarial settings who post content via anonymous accounts. The presence of extremist groups on the Internet is to achieve a specific aim or objective. Some of these objectives include communication, fundraising, sharing ideology, propaganda and community formation. Content analysis can examine the content to understand the aim behind the presence of extremist groups on the Internet. Content analysis can also help examine the behavioral patterns in the usage of these extremist websites. This can help understand the level of technical sophistication and interactivity quotient used by the owners of these websites. Various other content information like IP address, message headers in forums and comments in blogs can help gain insights into the way extremist users use and interact with each other. Content analysis can also investigate into the topics talked about by users and find virtual topic based communities.

\subsubsection{Modalities}

Various modalities are used to host and spread information on the Internet like websites, blogs and forums. Each of these modalities provide unique characteristics. For example, blogs provide an explicit option to add friends in the form of subscriptions. This information can be used to construct networks for further analysis. However, these modalities also bring a set of associated issues. For example, websites and forums may contain different types of content (text and/or multimedia) making it difficult to analyze the website. Also, nature of certain media like forums may not contain explicit links between users and forums even though they exist.

\subsubsection{Language}

Language is an important facet similar to \ref{subsubsec:language}. Language particularly creates problems for content based analysis techniques especially when multi-lingual content is analyzed.

\subsubsection{Genre}

Different genres of extremism are considered as a fact similar to \ref{subsubsec:genre}. This facet gives an understanding to which genres of extremism are analyzed.

\section{Survey of Automated Solutions for Online Radicalization Detection }
\label{sec:survey-solutions-detection}

In this section, we perform a survey of the techniques used to automatically detect online radicalization.  We present a summary of these solutions in Table  \ref{tab:detect-solutions-summary}.

\begin{sidewaystable}[htbp]
  \centering
   \tabcolsep=0.1cm
  \caption{Facets chosen for study of techniques to detect radical online content. Each facet has associated property and a brief description.}
    \begin{tabular}{ccll}
    \toprule
    \textbf{Facets} & \textbf{Label} & \textbf{Property} & \textbf{Description} \\
    \midrule
    \multirow{3}[0]{*}{\textbf{Technique}} & T1    & Link Based Bootstrapping (LBB) & Seed based bootstrapping on link structure \\
          & T2    & Text Classification & Radical content detection as binary text classification problem \\
          & T3    & Other & Other techniques apart from T1 and T2\\
    \midrule
    \multirow{4}[0]{*}{\textbf{Modality}} & M1    & Websites & Websites on the Internet \\
          & M2    & Online Forums & Privately and publicly accessible forums on the web \\
          & M3    & Blogs & Weblogs or blogs which are online equivalent of personal diaries \\
          & M4    & Social Media & Online social networking websites like Facebook, Twitter, YouTube etc. \\
    \midrule
    \multirow{3}[0]{*}{\textbf{Data Type}} & D1    & Text  & Documents consisting of only textual content (may include markups) \\
          & D2    & Multimedia & Documents containing multimedia content like audio, video and images \\
          & D3    & Text + Multimedia & Documents containing both text and multimedia \\
    \midrule
    \multirow{3}[0]{*}{\textbf{Features}} & F1    & Link   &  Features based on links between documents \\
          & F2    & Content & Features based on only content within the documents\\
          & F3    & Link + Content & Features based on both link and content (F1 and F2) \\
    \midrule
    \multirow{4}[0]{*}{\textbf{Evaluation}} & E1    & Precision & Evaluates the exactness of the technique \\
          & E2    & Recall & Evaluates the completeness of the the technique \\
          & E3    & F1 Measure & Weighted harmonic mean of Precision(E1) and Recall(E2) \\
          & E4    & Accuracy & Evaluates the correctness of the technique \\
    \midrule
    \multirow{3}[0]{*}{\textbf{Output}} & O1    & Users & Individuals using the Internet  \\
          & O2    & Content & Material on the Internet including text, audio, images and multimedia \\
          & O3    & Communities & Structural pattern of users based on interaction, social links etc. \\
    \midrule
    \multirow{3}[0]{*}{\textbf{Language}} & L1    & English & Documents consisting only of English content \\
          & L2    & Arabic & Documents consisting only of Arabic content \\
          & L3    & Multilingual & Documents consisting content in multiple languages \\
    \midrule
    \multirow{4}[0]{*}{\textbf{Genre}} & G1    & US Domestic & Radicalization originating from domestic US issues \\
          & G2    & Middle East & Radicalization originating from Middle East groups \\
          & G3    & Latin America & Radicalization originating from Latin America \\
          & G4    & Other & Radicalization originating or targeted against other groups and societies \\
    \bottomrule
    \end{tabular}%
  \label{tab:facets-detection}%
\end{sidewaystable}%

\begin{sidewaystable}[htbp]
  \centering
  \footnotesize
  \caption{Summary of Online Radicalization Detection Literature}
  \tabcolsep=0.1cm
    \begin{tabular}{p{1.3cm}|lcc|cccc|ccc|ccc|cccc|ccc|ccc|cccc}
    \toprule
          & \multicolumn{3}{c}{Technique} & \multicolumn{4}{c}{Modality}  & \multicolumn{3}{c}{Data Type} & \multicolumn{3}{c}{Features} & \multicolumn{4}{c}{Evaluation} & \multicolumn{3}{c}{Output} & \multicolumn{3}{c}{Language} & \multicolumn{4}{c}{Genre}      \\
       & T1    & T2    & T3    & M1    & M2    & M3    & M4    & D1    & D2    & D3    & F1    & F2    & F3    & E1    & E2    & E3    & E4    & O1    & O2    & O3    & L1    & L2    & L3    & G1    & G2    & G3    & G4 \\
           \midrule
    Greevy \& Smeaton 2004
   &      & \checkmark      &       &    \checkmark   &      &       &       &  \checkmark     &       &       &     & \checkmark      &       &    \checkmark   & \checkmark     &  \checkmark    & \checkmark     &       &  \checkmark     &       &  \checkmark     &       &     &       &       &       &  \\
   \bottomrule
   Aknine et al. 2005   &      &       & \checkmark     & \checkmark      &     &       &       & \checkmark     &       &       &      &     \checkmark  &       &       & \checkmark     &       &       &       & \checkmark     &       &       &       & \checkmark   &       &     &       &  \checkmark \\
   
  Reid et al. 2005 & \checkmark  	&       &       &  \checkmark    &      &       &       &      &       &  \checkmark    & \checkmark     &       &       &       &      &       &       &       &     & \checkmark      &       & \checkmark      &     &       &  \checkmark   &       &  \\
  
  Zhou et al. 2005  & \checkmark     &       &       & \checkmark      &      &       &       &      &       & \checkmark      & \checkmark    &       &       &       &      &       &       &       & \checkmark     &  \checkmark     &       &       &      &   \checkmark    &      &       &  \\
  \bottomrule
  Last et al. 2006   &      & \checkmark      &       & \checkmark      &      &       &       & \checkmark    &       &       &     & \checkmark      &       &       &      &       &  \checkmark     &       & \checkmark     &       &       &       & \checkmark     &       &     &       &  \\
  \bottomrule
  Qin et al. 2007 & \checkmark    &       &       &   \checkmark    &      &       &       &      &       &  \checkmark     & \checkmark     &       &       &     &      &       &       &      & \checkmark     &       &       &       & \checkmark   &       & \checkmark     &       &  \\
  
 Zhou et al. 2007  & \checkmark     &       &       &       & \checkmark    &       &       &     &       &  \checkmark     & \checkmark     &       &       &       &      &       &       &       & \checkmark     &       &       &       & \checkmark    & \checkmark      &      &       &  \\
 \bottomrule
 Chen and Qin et al. 2008  & \checkmark    &       &       &   \checkmark    &      &       &       &      &       &  \checkmark     & \checkmark     &       &       &     &      &       &       &      & \checkmark     &       &       &       & \checkmark   &    \checkmark   & \checkmark     &   \checkmark    &  \\
 
 Chen and Chung et al. 2008  & \checkmark     &       &       & \checkmark      &      &       &       &      &       & \checkmark      & \checkmark     &       &       &       &      &       &       &       & \checkmark     & \checkmark      &       &       & \checkmark    &       & \checkmark     &       &  \\
 \bottomrule
Fu et al. 2010 & \checkmark     &       &       &       & \checkmark     &       &       &      &       &  \checkmark     & \checkmark     &       &       & \checkmark      & \checkmark     & \checkmark   &       &       &  \checkmark     &  \checkmark     &       &       &     \checkmark  &  \checkmark     & \checkmark     & \checkmark      &  \\

Huang et al. 2010 &    & \checkmark      &       &       &      &       &   \checkmark    &    &       &  \checkmark     &      & 
\checkmark      &       &        &      &       & \checkmark      &       & \checkmark     &       &  \checkmark     &       &     &     &    &       &  \checkmark \\

Sureka et al. 2010 &      &       &  \checkmark     &       &     &       &  \checkmark   &      & \checkmark    &       & \checkmark     &       &       &  \checkmark     &      &       &       &  \checkmark     & \checkmark    & \checkmark      &       &       &     &       &    &       & \checkmark  \\
\bottomrule
Qin et al. 2011 & \checkmark     &       &       &   \checkmark    &      &       &       &      &       &  \checkmark     & \checkmark     &       &       &      &     &       &       &       & \checkmark     &       &       &       &  \checkmark  &    \checkmark   & \checkmark   &   \checkmark    &  \\

    \bottomrule
    \end{tabular}%
  \label{tab:detect-solutions-summary}%
\end{sidewaystable}

\subsection{Link Based Bootstrapping (LBB) Techniques for Detecting Online Radicalization}
\label{sec:survey-webmining-detection}

Link Based Bootstrapping (LBB) techniques are the amongst the most popular techniques used to detect online radical content \cite{Reid_Qin_Zhou_Lai_Sageman_Weimann_Chen_2005} \cite{1511999} \cite{DBLP:journals/ijmms/QinZRLC07} \cite{4076513} \cite{237266} \cite{ASI:ASI20838} \cite{Qin:2011:MES:1968924.1968970}. These techniques use a semi-automated approach to detect radical content on various modalities on the Internet including websites, blogs and online forums. First, a set of seed URLs are identified from authoritative sources. The URLs are then expanded using back link search and favorite links to accumulate related URLs. The idea is that extremist websites (or forums) link to each other and form some sort of a community structure. The expanded set is once again manually filtered by domain experts in order to avoid collecting off-topic pages. Web crawlers are used to download and collect content. Extremist forums may not be indexed my modern search engines forming a part of the web named the ``Hidden Web". Not only do these forums consist of rich multimedia content but also face accessibility issues like membership and adversarial detection.  Hence, a semi-automated focused incremental crawler in conjunction with a recall-improvement based incremental update procedure is used to collect extremist forums on the ``Hidden Web" \cite{ASI:ASI21323} . \emph{Focused crawlers seeks, acquires, indexes, and maintains pages on a specific set of topics that represent a relatively narrow segment of the web}\cite{781636}. Forums which do not allow anonymous access require human expertise to apply for memberships to the webmaster. Appropriate spidering parameters, proxies, URL ordering techniques and wrapper parameters are chosen to make sure that maximum data is indexed also ensuring incremental updates. Duplicate content removal techniques are used to avoid multiple indexing. This approach yields better results than standard spidering in conjunction with periodic or incremental update procedures. The Dark Web portal project heavily employs LBB techniques to collect extremist data \cite{springerlink:10.1007/978-3-540-25952-7_10} \cite{DBLP:conf/isi/QinZLRSC05}. This data collection includes multiple modalities (like websites and online forums) and  various genres of hate like jihadist and US domestic hate.\footnote{\url{http://128.196.40.222:8080/CRI_Indexed_new/login.jsp}}  

Sureka et al. also use LBB technique to identify extremist videos, users and implicit communities on YouTube \cite{Sureka_Kumaraguru_Goyal_2010}. In particular, they manually identified a set of seed videos and then exploit the social relationships like friends, comments, subscriptions, favorites and playlists to expand the initial seed data. Conservative expansion is performed, in order to avoid off topic videos, based on in-degree, hub, information, out-degree and betweenness centrality. Starting from a seed of 60 users they were able to add 98 new users with 88\% average precision.

%The major of drawback of web mining techniques used are the manual interventions at almost each step of the process. These techniques do reduce the arduousness from manual searches but they are still far from the ideal. 

\subsection{Text Classification Techniques for Detecting Online Radicalization}
\label{sec:survey-textclassification-detection}

Greevy and Smeaton treat the problem of detecting racist texts as binary classification problem \cite{Greevy:2004:CRT:1008992.1009074}. They train a machine learning classifier, Support Vector Machine (SVM), and use Bag-of-Words (BOW), Bigrams and Part-of-Speech (POS) feature representations. They report best results for SVM with a polynomial kernel function and BOW features. They observe that a polynomial kernel function with the BOW representation performed outperforms other combinations. 

Last et al. propose a machine learning approach based on graph representation of web documents to classify multi-lingual terrorist content \cite{DBLP:conf/wisi/LastMK06}. They exploit the structure of the web page (title, anchor text and body) to create graph based representation of the documents. They use a sub-graph extraction algorithm in conjunction with a sub graph frequency-- inverse sub graph frequency metric to build the final representation. They used a C4.5 classifier to classify 648 Arabic web documents in terrorist or non-terrorist.

Fu et al. and Huang et al. use a machine learning approach on user-generated content to classify extremist videos on video sharing websites \cite{5137295} \cite{ASI:ASI21291}. They create a testbed of 224 positive and negative videos on YouTube, manually tagged by extremism experts, to evaluate their approach. The user-generated content includes description, title, author names, names of other uploaded videos, author name, comments, categories and tags. They used four feature sets : lexical(character-based, word-based), syntactic(frequency of function and punctuation words), content-specific features(word and character n-grams, video tags and categories) and feature selection strategy (using Information Gain) as training data for their classifiers. They used three classifiers: C4.5, Naive Bayes and SVM to train and test their model. The report best accuracy results on feature selection strategy document representation with SVM. Lexical and Syntactic features are reported as the key discriminating features. 

\subsection{Other Techniques for Detecting Online Radicalization}
\label{sec:survey-other-detection}

There are other techniques explored too for detecting online radicalization. Aknine et al. treat the problem of racist text classification as a 3-class problem : racist, anti-racist and neutral. They use a multi-agent based  approach to detect racist documents on the Internet \cite{springerlink:10.1007/11577935_17}. They use three agents : query agents(to query the web),  document agents(to fetch documents) and criteria agents(for linguistic features) in a pyramidal co-ordination framework. They evaluate their system on English, French and German languages.

\section{Survey of  Solutions for Online Radicalization Analysis }
\label{sec:survey-solutions-analysis}

In this section, we perform a survey of the techniques used to analyze online radicalization. We present a summary of these solutions in Table \ref{tab:analysis-summary}.

\begin{sidewaystable}[htbp]
  \centering
   \tabcolsep=0.26cm
  \caption{Facets chosen for study of techniques to analyze radical online content. Each facet has associated property and a brief description.}
    \begin{tabular}{lp{0.5cm}lll}
    \toprule
    \textbf{Facets} &    \textbf{Label} &     \textbf{Property} &     \textbf{Description} \\
    \midrule
    \multicolumn{1}{c}{\multirow{3}[0]{*}{\textbf{Type of Analysis}}} & A1  & Network  & Analysis of the topological and social links \\
    \multicolumn{1}{c}{} & A2    & Content & Analysis of the content (structural, textual)  \\
    \multicolumn{1}{c}{} & A3    & Network + Content & Analysis of both the network and content \\
    \midrule
    \multicolumn{1}{c}{\multirow{3}[0]{*}{\textbf{Modalities}}} & M1    & Websites & Websites on the Internet \\
    \multicolumn{1}{c}{} & M2    & Online Forums & Privately and publicly accessible online forums \\
    \multicolumn{1}{c}{} & M3    & Blogs & Weblogs or blogs which are online equivalent of personal diaries \\
    \midrule
    \multicolumn{1}{c}{\multirow{4}[0]{*}{\textbf{Network Analysis}}} & N1    & Communities & Group of nodes with explicit or implicit common interests \\
    \multicolumn{1}{c}{} & N2    & Central Leaders & Key players in the network on basis of information flow, popularity \\
    \multicolumn{1}{c}{} & N3    & Topology & Topological characteristics like Clustering Coefficient, Degree Distribution\\
    \midrule
    \multicolumn{1}{c}{\multirow{5}[0]{*}{\textbf{Content Analysis}}} & C1   & Website Activities &  Propaganda, fundraising, recruitment etc. \\
    \multicolumn{1}{c}{} & C2    & Authorship Analysis & Analyzing the ownership of posted content \\
    \multicolumn{1}{c}{} & C3    & Affect Analysis & Analyzing the emotions expressed on various topics \\
    \multicolumn{1}{c}{} & C4    & Internet usage & Analyzing the technical sophistication and interactivity of the content \\
    \multicolumn{1}{c}{} & C5    & Others & 	Other analysis including anomaly detection, topic detection \\
    \midrule
    \multicolumn{1}{c}{\multirow{3}[0]{*}{\textbf{Language}}} & L1    & English & Documents consisting only of English content \\
    \multicolumn{1}{c}{} & L2    & Arabic & Documents consisting only of Arabic content \\
    \multicolumn{1}{c}{} & L3    & Multilingual & Documents consisting content in multiple languages \\
    \midrule
    \multicolumn{1}{c}{\multirow{4}[0]{*}{\textbf{Genre}}} & G1    & US Domestic &  Radicalization originating from domestic US issues \\
    \multicolumn{1}{c}{} & G2    & Middle East & Radicalization originating from Middle East groups \\
    \multicolumn{1}{c}{} & G3    & Latin America & Radicalization originating from Latin America \\
    \multicolumn{1}{c}{} & G4    & Others &  Radicalization originating or targeted against other groups and societies \\
    \bottomrule
  \label{tab:facets-analysis}%
  \end{tabular}%
\end{sidewaystable}%

% Table generated by Excel2LaTeX from sheet 'Sheet2'
\begin{sidewaystable}[htbp]
  \centering
  \tabcolsep=0.15cm
  \caption{Summary of Online Radicalization Analysis Literature}
    \begin{tabular}{p{1.3cm}ccc|ccc|ccc|ccccc|ccc|ccccc}
    \toprule
   & \multicolumn{3}{c}{\textbf{Analysis}} & \multicolumn{3}{c}{\textbf{Modalities}} & \multicolumn{3}{c}{\textbf{Network}} & \multicolumn{5}{c}{\textbf{Content}  }& \multicolumn{3}{c}{\textbf{Language}} & \multicolumn{4}{c}{\textbf{Genre}}     & \\
    \midrule
    \multicolumn{1}{c}{} & A1    & A2    & A3    & M1    & M2    & M3    & N1    & N2    & N3    & C1    & C2    & C3    & C4    & C5    & L1    & L2    & L3    & G1    & G2    & G3    & G4 \\
    \multicolumn{1}{l}{Abbasi and Chen 2005} &     & \checkmark    &     &     & \checkmark    &     &     &     &     &     & \checkmark    &     &     &     & \checkmark    & \checkmark    &     &     & \checkmark    &     &  \\
    
    \multicolumn{1}{l}{Reid et al. 2005} &     &     & \checkmark    & \checkmark    &     &     & \checkmark    &     &     & \checkmark    &     &     &     &     &     &     & \checkmark    &     & \checkmark    &     &  \\
        
    \multicolumn{1}{l}{Zhou et al. 2005} &     &     & \checkmark    & \checkmark    &     &    & \checkmark    &     &     & \checkmark    &     &     &     &     & \checkmark    &     &     & \checkmark    &     &     &  \\
            
      \bottomrule        
     
    \multicolumn{1}{l}{Xu et al. 2006} & \checkmark    &     &     & \checkmark    &     &     &     &     & \checkmark    &     &     &     &     &     &     &     &     & \checkmark    & \checkmark    & \checkmark    &  \\
    
    \bottomrule

    \multicolumn{1}{l}{Abbasi et al. 2007} &     & \checkmark    &     &     & \checkmark    &     &     &     &     &     &     & \checkmark    &     &     & \checkmark    & \checkmark    &     & \checkmark    & \checkmark    &     &  \\

         \multicolumn{1}{l}{Chau and Xu 2007} & \checkmark    &     &     &     &     & \checkmark    & \checkmark    & \checkmark    & \checkmark    &     &     &     &     &    & \checkmark    &     &     & \checkmark    &     &     &  \\                        

   \multicolumn{1}{l}{Chen et al. 2007} &     &     & \checkmark    & \checkmark    &     &     & \checkmark    &     &     & \checkmark    & \checkmark    &     &     &     & \checkmark    & \checkmark    &   &     &     &     &  \\

         \multicolumn{1}{l}{Qin et al. 2007} &     & \checkmark    &     & \checkmark    &     &     &     &     &     &     &     &     & \checkmark    &     &     &     & \checkmark    &     & \checkmark    &     &  \\
         
    \bottomrule

         \multicolumn{1}{l}{Chen 2008} &     & \checkmark    &     &     & \checkmark    &     &     &     &     &     &     & \checkmark    &     &     &     & \checkmark    &     &     & \checkmark    &     &  \\    

         \multicolumn{1}{l}{Chen et al. 2008} &     &     & \checkmark    & \checkmark    &     &     & \checkmark    &     &     & \checkmark    &     &     &     &     &     & \checkmark    &     &     & \checkmark    &     &  \\

         \multicolumn{1}{l}{Fu et al. 2008} & \checkmark    &     &     & \checkmark    &     &     & \checkmark    &     & \checkmark    &     &     &    &     &    &     &     &     &     &     &     & \checkmark \\

         \multicolumn{1}{l}{Mielke et al. 2008} &     & \checkmark    &     & \checkmark    &     &     &     &     &     &     &     &     &     & \checkmark    &     & \checkmark    &     &     & \checkmark    &     &  \\

         \multicolumn{1}{l}{Qin et al. 2008} &     & \checkmark    &     & \checkmark    &     &     &     &     &     &     &     &     & \checkmark    &     &     &     &     & \checkmark    & \checkmark    & \checkmark    &  \\

         \multicolumn{1}{l}{Xu et al. 2008} & \checkmark    &     &     & \checkmark    &     &     &     &     & \checkmark    &     &     &     &     &     &     &     & \checkmark    &     & \checkmark    &     & \checkmark \\

    \bottomrule

         \multicolumn{1}{l}{Yang et al. 2009} &     & \checkmark    &     & \checkmark    &     &     &     &     &     &     &     &     &     & \checkmark    &  \checkmark   &     &     &     & \checkmark    &     &  \\
         
          \multicolumn{1}{l}{Zhang et al. 2009} &     & \checkmark    &     &     & \checkmark    &     & \checkmark    &     &     &     &     &     &     & \checkmark    &     &     & \checkmark    &     &     \checkmark &     &  \\
         
	\bottomrule    

           \multicolumn{1}{l}{Kramer 2010} &     & \checkmark    &     &     & \checkmark    &     &     &     &     &     &     &     &     & \checkmark    &     &    & \checkmark    &     & \checkmark    &     &  \\

                        \multicolumn{1}{l}{L`Huillier et al. 2010} &     &     & \checkmark    &     & \checkmark    &     & \checkmark    & \checkmark    &     &     &     &     &     & \checkmark    & \checkmark    &     &     &     & \checkmark    &     &  \\
                                           
	\bottomrule
	
                        \multicolumn{1}{l}{Qin et al. 2011} &     & \checkmark    &     & \checkmark    &     &     &     &     &     &     &     &     & \checkmark    &     &     &     & \checkmark    & \checkmark    & \checkmark    & \checkmark    &  \\
            
    \bottomrule
    \end{tabular}%
  \label{tab:analysis-summary}%
\end{sidewaystable}%

\subsection{Network Based Analysis for Online Radical Content}

Link based analysis exploits the structure of links between nodes in a network to analyze the topological characteristics of a network. Websites on the Internet contain hyperlinks which refer to external as well as internal web pages. These hyperlinks can be analyzed to gain further insight into the network's topological characteristics. 

Various approaches in literature analyze such a hyperlink structure between extremist websites \cite{Reid_Qin_Zhou_Lai_Sageman_Weimann_Chen_2005} \cite{1511999} \cite{springerlink:10.1007/978-3-540-71549-8_1} \cite{ASI:ASI20838} \cite{5137274}. Typically, the entire extremist network is treated as a graph where each extremist website is a node and the edges are hyperlinks. One of the main objectives of exploring links between websites is to detect implicit communities and visualize them. In order to achieve that objective, each link between the nodes is assigned a weight depending on the page level in the site hierarchy. This graph is then fed to a visualization algorithm, Multi-Dimensional Scaling (MDS). MDS arranges the nodes in the network such that the most similar nodes are close to each other while the non-similar nodes are far off from each other.  Many studies use MDS along with the page level similarity metric to discover hyper linked extremist communities \cite{Reid_Qin_Zhou_Lai_Sageman_Weimann_Chen_2005} \cite{1511999} \cite{springerlink:10.1007/978-3-540-71549-8_1} \cite{ASI:ASI20838}. In case of blogs sharing websites, each node is treated as a user and an edge is either a subscription or a group co-membership (belonging to the same blog ring) relationship \cite{Chau200757}. Similarly, in case of forums each user is marked as a node and a reply/message on a thread (interaction/activity) is treated as a link between the two users. In addition to finding communities, it's also important to find central players in these communities.  Zhou et al. find central nodes in a collection of US domestic hate groups \cite{1511999}. Chau et al. use betweenness and degree measures to discover  central nodes in White Supremacist blogs in the US \cite{Chau200757}. Central nodes help structure other nodes and facilitate information flow in a network. These central nodes can be vital break down points in order to debilitate such networks.

Apart from analyzing the link structure in a networks, it's also important to investigate the topological characteristics of networks. Average Path Length, Clustering Coefficient, Degree distribution are a few metrics widely used to explain network topologies. These metrics are usually calculated on the giant component in a network. Giant component is the largest connected (not disjoint) component in a graph and hence, a representative subset of the network in study. Chau et al. analyze the topological characteristics of the giant component in anti-Black extremist blog rings on Xanga, a popular blog sharing website  \cite{Chau200757}.  Xu et al. analyze the structural characteristics of Middle-Eastern, Latin American, US domestic extremist websites, Global Salafi Jihad (terrorist) network and Meth Word network (illegal drug trafficking network) \cite{DBLP:conf/isi/XuCZQ06} \cite{Xu:2008:TDN:1400181.1400198}. In all the above studies, network characteristics like average path length, clustering coefficient and degree distribution were calculated. The degree distribution follow a power-law and highlights preferential attachment treatment in these networks where new users are attracted to popular nodes. Such networks are called scale-free networks and depict the ``rich get richer" phenomena. The average path length observed is small indicating these networks are relatively small-world. This shows that a node can connect with any other node in very few hops (five or less). A high clustering coefficient shows that there are dense local links leading to optimal efficiency in communication between nodes.

In some cases, just studying the topology of the network isn't enough. A combination of both network and content is considered desirable to enable deep understanding of the network in study. L'Hullier et al propose a topic centered social network analysis approach to discover virtual communities in online extremist forums \cite{LHuillier:2010:TSN:1938606.1938615}. They first create a network with links which show interaction between members in the forums. A link between two members signifies interaction between the members like posting a message on a common thread. Members are then filtered according to topics to find topic centered virtual communities. These topics are extracted using Latent Dirichlet Allocation (LDA), a topic modeling algorithm. HITS algorithm is employed to find information hubs and authoritative members in the network.

\subsection{Content Based Analysis for Online Radical Content}

Content based analysis investigates patterns in the content posted by extremist websites and groups. These patterns can lead to insights in the functioning of these extremist organizations and the purpose of the content posted by these groups. 

Users in extremist forums may post hate promoting content under the condition of anonymity. Authorship analysis can help attributing content to the actual owner in an adversarial setting. Authorship analysis profiles the stylistic variations in the content posted by a user. Some studies apply authorship analysis to multi-lingual (English and Arabic) messages in online extremist forums \cite{1512002} \cite{springerlink:10.1007/978-3-540-71549-8_1}. Authorship identification is treated as a multi-class text classification problem where each class is an author. A combination of multiple features including lexical (world length distribution, frequency of letters), syntactic (punctuation, function words), structural (font size, font color) \& content-specific (race, gender) are used to profile user's content and document representation. In order to address, Arabic language specific issues like diacritics, inflection and word elongation an Arabic language parser is used. Two classifiers C4.5 and SVM are used and feature sets are added incrementally. SVM with combination of all feature sets perform the best on both English and Arabic messages.

Extremist groups use and maintain websites for various purposes to achieve their objective of xenophobia. A few studies manually classify extremist websites into eight pre-defined categories based on their content \cite{Reid_Qin_Zhou_Lai_Sageman_Weimann_Chen_2005}  \cite{1511999}  \cite{springerlink:10.1007/978-3-540-71549-8_1} \cite{ASI:ASI20838}. These categories include communications, fundraising, sharing ideology, propaganda(insiders), propaganda (outsiders), virtual community, commands \& control and recruitment \& training. The strength (or intensity) of these categories are displayed using Snowflake Visualization. Snowflake visualization helps display the intensity of a variable across multiple dimensions \cite{Reid_Qin_Zhou_Lai_Sageman_Weimann_Chen_2005} \cite{ASI:ASI20838}  \cite{Chau200757}. These studies show that most terrorist organizations primarily use the Internet to share their ideology.

Affect analysis measures the quantity of affect (or emotion) in given content. Affect analysis can help understand the emotions of users towards various topics. Few studies analyze the intensity of emotions in international extremist forums \cite{4258712}   \cite{4565038}.  Abbasi manually creates an affect lexicon using a probability disambiguation technique \cite{4258712}. The affect lexicon is used to calculate an affect intensity score for each class. Abbasi reports that the intensity of hate and violence related affects are higher in Middle Eastern forums than US domestic extremist forums. Chen uses a machine learning classifier to analyze affects in two jihadist web forums : Al Firdaws 	and Montada   \cite{4565038}. Various linguistic features like character n-grams, word n-grams, root n-grams and collocations are extracted as indicators. A recursive feature elimination technique (RFE) in conjunction with Information Gain (IG) heuristic is used to reduce feature dimensions. An SVR ensemble classifier is trained with and evaluated on a manually created test bed. Al Firdaws postings contained more violence, hate and racism related affects than messages on Al Montana forum.

Few studies analyze extremist websites in terms of its technical sophistication, content richness and web interactivity  \cite{DBLP:journals/ijmms/QinZRLC07}   \cite{237266}   \cite{Qin:2011:MES:1968924.1968970}. Technical sophistication attributes include use of dynamic web programming, embedded multimedia, dynamic web programming scripts. Content Richness attributes include hyperlinks, software and file downloads. Web interactivity attributes include guest books, chat rooms, online shops and feedback forms. Studies report that extremist websites make use of advanced communication methods like email and chat to facilitate flow of information. These studies also show that extremist groups also more of multimedia content (images, video) in order to make a quick and impressionable impact on the eye of the reader.

 \subsection{Other Solutions for Analyzing Online Radical Content}

There are some other types of analysis performed in extremist online forums. Some studies also analyze the content to discover important topics in extremist content \cite{4925106}. A topic modeling algorithm like Latent Dirichlet Allocation (LDA) is used to model the documents as a collection of topics. Other studies map the geo locations of ISPs, cities and countries of Jihadist extremist web sites \cite{springerlink:10.1007/978-3-540-89900-6_12}. The mapping reveals that major extremist websites are hosted in the US or Europe based countries. Kramer  proposes to use Finite Lyapunov Time Exponents (FLTE) to find anomalies in extremist forums postings \cite{Kramer:2010:ADE:1938606.1938614}. FLTEs are considered good indicators of stability in dynamic systems and have wide applications in control systems research.  The idea is that the distribution of words in an anomalous environment changes from the normal distribution. Kramer models the text posted in forums by each user as a time-series equation. tf-idf scores of the words posted by the user are used to form equations. FLTEs are then used to check the ``change" in distributions which if positive amounts to an anomaly.

\section{Discussion}
\label{sec:discussion}

In this section, we discuss the analysis of facets selected for paper classification with respect to the initial objectives of this literature survey. We revisit the solution space and synthesize the solutions to determine trends and limitations.

\subsection{Online Radicalization Detection}

\begin{figure*}
\centering
\includegraphics[angle= 90, scale = 0.6] {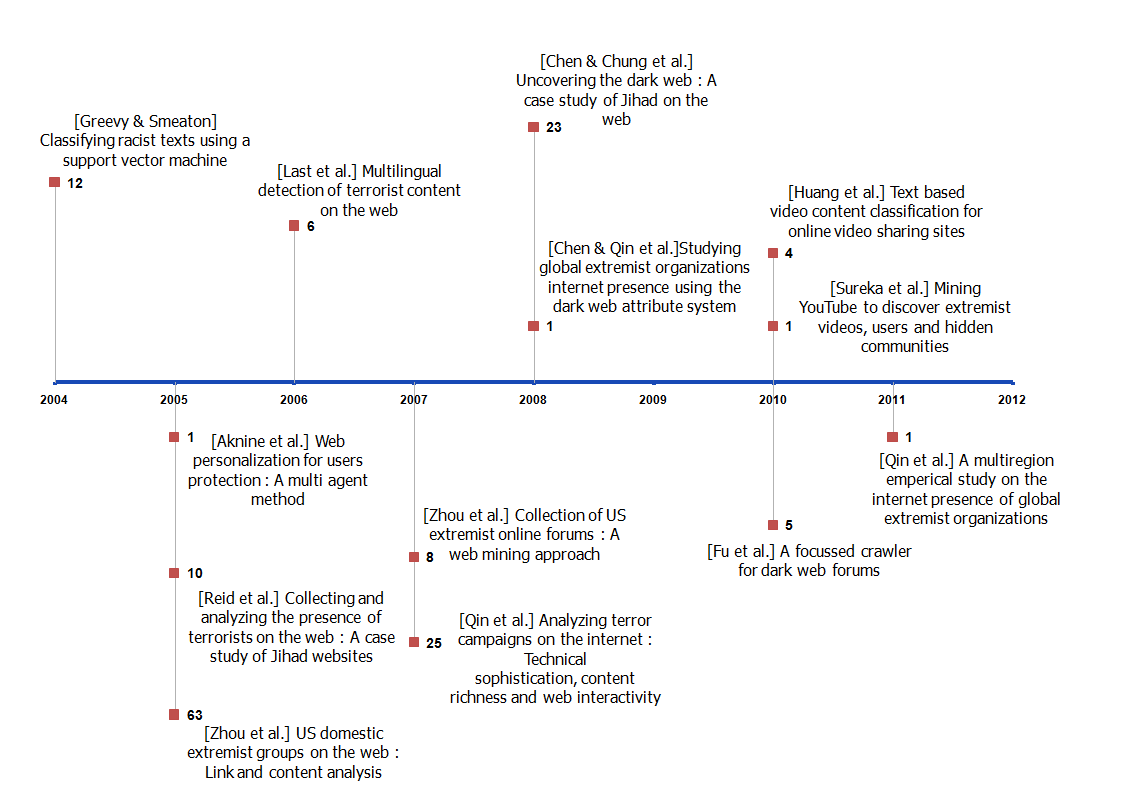}
\caption{Timeline of literature in Online Radicalization Detection. The red square box denotes the number of citations received. The distance from the axis is proportional to the number of citations received by each paper.}
\label{fig:timeline-detection}
\end{figure*}

Figure \ref{fig:timeline-detection} shows the timeline of literature in online radicalization detection. Each paper is depicted by a red square box accompanied by the author and the title of the paper. The distance from the axis is proportional to the number of citations currently received by the paper.

\begin{figure*}
\includegraphics[angle=90, scale=0.5]{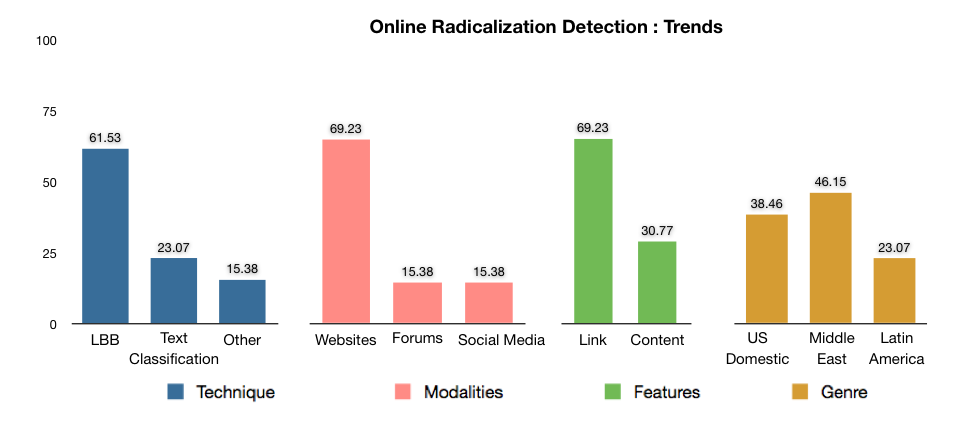}
\centering
\caption{Trend Analysis of literature in online radicalization detection. The percentages show the number of papers included in our survey which contain the corresponding dimension}
\label{fig:detection} 
\end{figure*}

\begin{enumerate}[(a)]
	
	\item \emph{Techniques}
	
	There are two major techniques used to detect (collect or search or find) online radical content : Link Based Bootstrapping (LBB) and Text Classification. LBB makes use of a startup or seed data (identified by extremist experts) and then exploits the link (typically hyperlink or social links) in the data to snowball the seed and obtain an expanded set. The irrelevant or off-topic data in the expanded set is filtered out manually by domain experts. Text classification on the other hand handles the problem of detecting online radical content as a binary text classification problem. It examines the linguistic features in the content and uses a machine learning classifier to make decisions.
	
	\item \emph{Trends}
	
	Figure \ref{fig:detection} shows the trends observed in literature in the solution space of online radicalization detection. We observe that LBB technique is the most popular technique to detect online radicalization. These techniques exploit the \emph{co-citation} phenomena where ``like minded" groups link to each other. Hence, most of these techniques use link based features rather than content based features. We also notice that the most of the literature detect extremist content on websites. Relatively less attention is devoted to online forums and social network websites. Amongst social media websites solutions are targeted towards YouTube, a popular video sharing website. We reason that this may be due to the appealing nature of multi-media and its immediate impact. Other popular online social media like Twitter, Facebook and Google+ are ignored. Also, Middle Eastern genre of extremism is the most popular followed closely by US domestic and Latin American hate.

	\item \emph{Limitations}
	
	Both LBB and text classification techniques proposed in literature have their limitations. LBB techniques are semi-automated and require tremendous manual effort to filter out irrelevant content. Content on the Internet is very ephemeral and tends to change patterns in time. LBB based techniques are difficult to cope up on temporal and fleeting extremist content. Text classification techniques on the other hand make some assumptions on the data. The task of detecting radical content is treated as a binary classification problem. However, the amount of radical content on the Internet is small than the amount of non-radical content. Hence, the dataset used to evaluate text classification should be an imbalanced dataset. This aspect is not being considered in solutions which use text classification to solve the problem of online radicalization \cite{springerlink:10.1007/11577935_17} \cite{Greevy:2004:CRT:1008992.1009074}.

\end{enumerate}

\subsection{Online Radicalization Analysis}

\begin{figure}
\centering
\includegraphics[angle= 90, scale = 0.6] {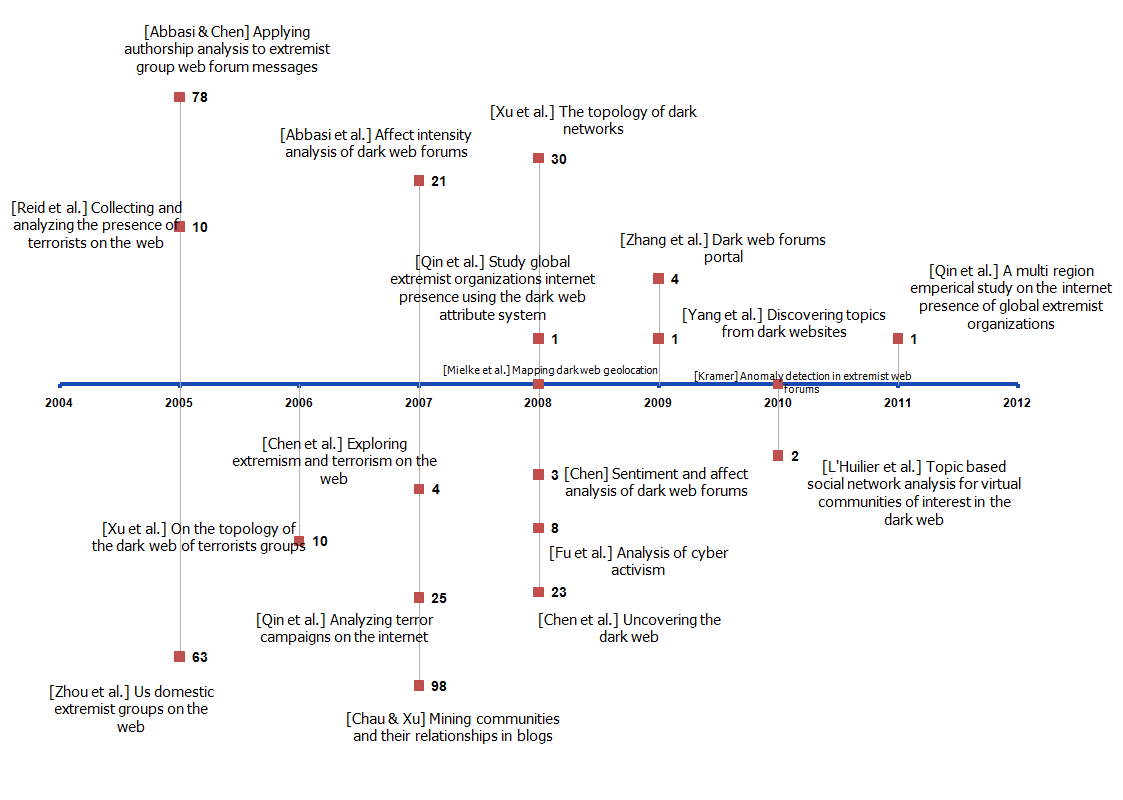}
\caption{Timeline of literature in the space Online Radicalization Analysis. The red square box denotes the number of citations received. The distance from the axis is proportional to the number of citations received by each paper.}
\label{fig:timeline-analysis}
\end{figure}

Figure \ref{fig:timeline-analysis} shows the timeline of literature in the space of online radicalization analysis. Each paper is depicted by a red square box accompanied by the author and the partial title of the paper (for better readability). The distance from the axis is proportional to the number of citations currently received by the paper.

\begin{figure*}
\includegraphics[angle=90, height=7in, width=4in]{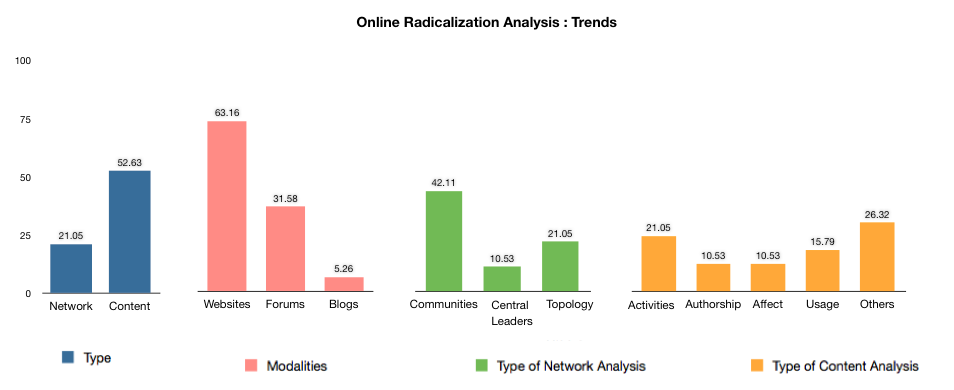}
\centering
\caption{Trend Analysis of literature in online radicalization analysis. The percentages show the number of papers included in our survey which contain the corresponding dimension}
\label{fig:analysis} 
\end{figure*}

\begin{enumerate}[(a)]
	
	\item \emph{Type of Analysis}
	
	Online radical content can be analyzed either in terms of the network formed (structure of hyperlinks or social links) or by the information it contains. The main goal of analyzing the content is to provide actionable information to law enforcement agencies. Various network based (communities, leaders and topological characteristics) and content based (authorship identification, website activities, affect analysis, usage) are performed on online radicalization content.
	
	\item \emph{Trends}
	
	Figure \ref{fig:analysis} shows the trends observed in literature in the solution space of online radicalization detection.  Network based analysis is the most popular type of analysis performed on online radical content. Most network analyzes focus on detecting communities and their characteristics as these are the most pertinent information to a law enforcement agency or security analyst. Also, the most common modalities focused on for analysis are websites closely followed by online forums. We argue that this may be because Middle-Eastern extremist groups (most studied genre of extremism) are technically sophisticated to host websites and online forums. These modalities allow these groups better control over users, content and activity. These modalities also allow the existence of various interactive methods like live broadcast radio and chat rooms. 
	
	\item \emph{Limitations}
	
	Different network based and content based techniques are used to analyze radical content. Network based techniques throw light on the community structure, key leaders and topological characteristics of the network. The techniques used in the literature don't use formal community detection algorithms to detect communities \cite{Reid_Qin_Zhou_Lai_Sageman_Weimann_Chen_2005} \cite{Chau200757} \cite{LHuillier:2010:TSN:1938606.1938615}. These techniques rely more on visual inspection using graph layouts which can be misleading and unreliable. Content based techniques analyze the structure \& type of content to gauge an understanding of the purpose of the extremist website like propaganda, fundraising, sharing ideology etc. Currently, the techniques in literature perform this classification manually by the help of extremism experts \cite{springerlink:10.1007/978-3-540-71549-8_1} \cite{237266}. In practice, this may not be a feasible solution for a security analyst. 
	
\end{enumerate}

\section{Research Gaps}
\label{sec:research-gaps}

Based on our classification scheme and its analysis of literature in the space of \emph{Automated Online Radicalization Detection and Analysis} we identify few research gaps based on the facets we considered for our paper classification scheme.

\begin{enumerate} [(I)]

	\item \emph{Modalities : Twitter, Facebook}
	
	Each modality (micro-blogging website, video-sharing website, photo-sharing website, blogosphere, online chat, online forums or discussions threads) poses unique technical challenges and hence a focused approach is required to develop techniques for a specific modality. While there has been work done in the area of video-sharing websites such as YouTube and blogosphere, micro-blogging website (Twitter being the most popular example) is a modality which is unexplored. We also notice that mining FaceBook in the context of automatic detection of online radicalization is an area which is relatively unexplored.
	
	\item \emph{Online Radicalization Detection : Activity Based Detection}
	
	Our survey reveals that content-based and link-based features have been used as effective signals to identify hate-promoting content. However, activity-based features (usage-based) are not yet explored. We believe that investigating the application of activity-based features for the task of hate-promoting content and user detection is an open research problem.

	\item \emph{Online Radicalization Analysis: Community Detection }
	
	Community formation is an important property of most social networks like WWW, Twitter and Facebook. Hence, they are significant actionable information for law enforcement agencies. In literature, we observe that community detection is performed by visual layouts. However, algorithms for community  detection have been well studied and evaluated on real world networks like WWW and collaboration networks. These algorithms haven't been applied on extremist networks. Moreover, networks on social media like Twitter are constantly evolving and temporal in nature. Therefore, there's a need of studying both static as well as dynamic community detection algorithms.

\end{enumerate}

\section{Conclusion}
\label{sec:conclusion}

Automated solutions to detect and counter cyber-crime (to address the needs of law-enforcement and intelligence-agencies) related to promotion of hate and radicalization on the Internet is an area which has recently attracted a lot of research attention. In this survey paper, we reviewed the state-of-the-art in the area of automated solutions for detecting online radicalization on Internet and social media websites. We presented a novel multi-level taxonomy or framework to organize the existing literature and present our perspective on the area. We categorized existing studies (based on a multi-level taxonomy) on various dimensions such as the social media websites (YouTube, Twitter, FaceBook), technique employed (Machine Learning, Social Network Analysis) and modality (Websites, Online Forums, Blogs, Social Media). We reviewed key results, compared and contrasted various approaches and offered our fresh perspective on trends based on the evidence derived from our extensive collection of literature in the area. We also identified research gaps in this line of research, discussed unsolved problems and present potential future work. 

\begin{received}
Received November 2011; accepted December 2011
\end{received}

\bibliographystyle{acmtrans}
\bibliography{compre}

\end{document}